\documentclass[pre,twocolumn,showpacs]{revtex4-1}

\usepackage{amssymb,amsmath}

\usepackage{color}

\newcommand\beq{\begin{equation}}
\newcommand\eeq{\end{equation}}
\newcommand\beqa{\begin{eqnarray}}
\newcommand\eeqa{\end{eqnarray}}
\newcommand{\dd}{\text{d}}
\newcommand{\nn}{\nonumber\\}

\newcommand{\al}{\alpha}

\usepackage{graphicx}% Include figure files

\begin{document}

%\begin{frontmatter}

\title{Stability of freely cooling granular mixtures at moderate densities}
\author{Vicente Garz\'o}
\email{vicenteg@unex.es} \homepage{http://www.eweb.unex.es/eweb/fisteor/vicente/}
\affiliation{Departamento de F\'{\i}sica and Instituto de Computaci\'on Cient\'{\i}fica Avanzada (ICCAEx), Universidad de Extremadura, E-06071 Badajoz, Spain}

\begin{abstract}

The formation of velocity vortices and density clusters is an intriguing phenomenon of freely cooling granular flows. In this work, the critical length scale $L_c$ for the onset of instability is determined via stability analysis of the linearized Navier-Stokes hydrodynamic equations of $d$-dimensional granular binary mixtures at moderate densities. In contrast to previous attempts, the analysis is not restricted to nearly elastic systems since it takes into account the nonlinear dependence of the transport coefficients and the cooling rate on the collisional dissipation. As expected from previous results obtained in the very dilute regime, linear stability shows $d-1$ transversal (shear) modes and a longitudinal (``heat'') mode to be unstable with respect to long enough wavelength excitations. The theoretical predictions also show that the origin of the instability is driven by the transversal component of the velocity field that becomes \emph{unstable} when the system length $L>L_c$. An explicit expression of $L_c$ is obtained in terms of the masses and diameters of the mixture, the composition, the volume fraction and the coefficients of restitution. Previous results derived in the limit of both mechanically equivalent particles and low-density mixtures are consistently recovered. Finally, a comparison with previous theoretical works which neglect the influence of dissipation on the transport coefficients shows quantitative discrepancies for strong dissipation.

%In addition, a previous comparison [Mitrano \emph{et al.}, Phys Rev E (2014);89:020201 (R)] with molecular dynamics (MD) simulations shows good agreement between MD and kinetic theory results.

\end{abstract}

\draft
\date{\today}
\maketitle

\section{Introduction}
\label{sec1}

Granular media are involved in many industrial and natural phenomena. In fact, it has been estimated that granular media is the second most used type of material in industry after water \cite{AFP13}. This is perhaps the main reason for which the study of granular matter has attracted the attention of physicists and engineers in the past few years. Although granular media form an extremely vast family constituted by grains of different sizes and shapes, all these systems share relevant features. In particular, when granular materials are externally excited (rapid flow conditions), they behave like a fluid. In this regime, binary collisions prevail and hence, kinetic theory may be considered as a quite useful tool to describe the kinetics and hydrodynamics of the system. The main difference with respect to ordinary or molecular fluids is that granular systems are constituted by macroscopic grains that collide inelastically so that the total energy decreases with time. In this context, a granular fluid can be considered as a \emph{complex} system that inherently is in a non-equilibrium state. In the case that the system is heated by an external driving force that compensates for the energy dissipated by collisions, a non-equilibrium \emph{steady} state is achieved. In these conditions, some attempts have been recently made to formulate a \emph{fluctuation-response} theorem based on the introduction of an effective temperature \cite{DG01,PBL02,SL04,G04,SBL06,G07,PBV07,MPRV08}. The generalization of the equilibrium fluctuation-response theorem to non-equilibrium states has been also confirmed in real experiments of intruders in driven granular fluids \cite{GPSV14}. Another interesting experiments in granular matter have studied the response of a sheared granular medium in a Couette geometry \cite{BPDPPZ08} and the behaviour of a freely rotating asymmetric probe immersed in a vibrated granular media \cite{PP11}.

On the other hand, although significant progresses have been made in the past on the understanding of granular flows, there are still important open challenges in the research of granular gases. One of the main reasons for which the theoretical description of these systems is quite intricate is that the number of relevant parameters needed to describe them is relatively large. This gives rise to a wide array of complexities that arise during the derivation of kinetic theory models. Thus, in order to gain some insight into the description of these systems in real conditions, one usually models a granular fluid as a system composed by \emph{smooth} hard spheres or disks with inelastic collisions. In this simplest model, the inelasticity of collisions is only accounted for by a (positive) \emph{constant} coefficient of normal restitution $\al \leq 1$ that only affects the translational degrees of freedom of grains. Nevertheless, in spite of the simplicity of the model, it has been shown as a reliable prototype to explain some of the physical mechanisms involved in granular flows, especially those directly related to collisional dissipation.

One of the most characteristic features of granular fluids is the spontaneous formation of velocity vortices and density clusters in freely cooling flows (homogeneous cooling state, HCS). The origin of this kind of instability is associated with the dissipative nature of collisions and is likely the most characteristic feature that makes granular flows so distinct from ordinary (elastic) fluids. Detected first by Goldhirsch and Zanetti \cite{GZ93} and McNamara \cite{M93} in computer simulations, the instabilities in a free granular fluid can be well described by a linear stability analysis of the Navier-Stokes hydrodynamic equations. This analysis provides a critical length $L_c$ so that the system becomes unstable when its linear size is larger than $L_c$. In the case of a monodisperse low-density granular gas, the dependence of $L_c$ on the coefficient of restitution obtained from the (inelastic) Boltzmann kinetic equation \cite{BDKS98,DB03} compares quite well with numerical results \cite{BRM98} obtained by using the direct simulation Monte Carlo (DSMC) method \cite{B94}. For higher densities, theoretical results for $L_c$ based on the (inelastic) Enskog equation \cite{G05} shows an excellent agreement with molecular dynamics (MD) simulations for a granular fluid at moderate density \cite{MDCPH11,MGHEH12}. The stability analysis reported in Ref.\ \cite{G05} extends to finite dissipation some previous attempts \cite{Mathieu1,Mathieu2} carried out in the context of the Enskog kinetic theory but neglecting any dependence of the pressure and the transport coefficients on inelasticity. Nevertheless, while the study of the stability of the HCS has been widely covered in the case of granular fluids, much less has been made in the important subject of granular mixtures (namely, systems composed by grains of different masses, diameters, composition).

Needless to say, the analysis of the stability of the HCS for polydiperse granular systems is much more complicated than for a single granular gas. Not only the number of transport coefficients involved in the determination of the critical size $L_c$ is higher than for a monodisperse gas but also they depend on more parameters, such as the set of coefficients of restitution characterizing the binary collisions between different species. Many of the early attempts \cite{JM89,Z95,AW98,AJ04} to obtain the Navier-Stokes coefficients of granular mixtures were performed by assuming the equipartition of granular energy. However, given that the lack of energy equipartition \cite{GD99bis} has been widely confirmed by computer simulations \cite{MG02,BT02,DHGD02,KT03,WJM03} and observed in real experiments of agitated mixtures \cite{WP02,FM02}, the hypothesis of energy equipartition can only be acceptable for nearly elastic systems. In fact, in those previous works \cite{JM89,Z95,AW98,AJ04} the forms of the transport coefficients are the same as those obtained for ordinary mixtures \cite{LCK83} and the influence of inelasticity is only considered in the presence of a sink term in the energy balance equation. A more rigorous derivation of linear transport for granular mixtures has been made by Garz\'o and Dufty \cite{GD02} in the  dilute regime and more recently by Garz\'o, Dufty and Hrenya \cite{GDH07,GHD07,MGH12} for moderate densities. In these works, given that nonequipartition effects on transport have been considered, the corresponding Navier-Stokes transport coefficients exhibit an intricate nonlinear dependence on the coefficients of restitution of the mixture. As for single granular gases, the theoretical results (which have been obtained in the so-called fist Sonine approximation) compare in general quite well with computer simulations \cite{R00,MG03,GM03,GM04,GV09,GV12} for conditions of practical interest, such as strong inelasticity.

The knowledge of the Navier-Stokes transport coefficients of granular mixtures opens the possibility of obtaining the critical length $L_c$ from the (linear) stability analysis of the hydrodynamic equations. For dilute systems, the theoretical predictions of kinetic theory \cite{GMD07,BR13} for $L_c$ has been shown to agree very well with the DSMC simulations of the Boltzmann equation \cite{BR13}. On the other hand, in spite of the explicit knowledge of the Enskog transport coefficients for a granular mixture \cite{GDH07,GHD07,MGH12}, I am not aware of any previous solution of the linearized hydrodynamic equations for moderately dense granular mixtures. The goal of this paper is to perform a linear stability analysis around the HCS in order to identify the conditions for stability as functions of the wave vector, the volume fraction, the dimensionality of the system $d$ and the parameters of the mixture (masses, sizes, composition and the coefficients of restitution). As expected, the stability analysis shows $d-1$ transversal (shear) modes and a longitudinal heat mode to be unstable with respect to long wavelength excitations. In addition, the results also show that the origin of the instability lies in the transversal shear mode (except for quite large dissipation) and hence, for sizes of the system larger than the critical length $L_c$ the transversal velocity becomes unstable. As for dilute mixtures \cite{BR13}, theoretical predictions for $L_c$ compare well with recent MD simulations of hard spheres \cite{MGH14}. A preliminary short report of some of the results presented here has been given in Ref.\ \cite{MGH14}.

The plan of the paper is as follows. First, in Section \ref{sec2} the hydrodynamic equations and associated fluxes to Navier-Stokes order are recalled. The explicit dependence of some of the transport coefficients on dissipation is illustrated for different systems showing that the influence of inelasticity on transport is in general quite significant. Section \ref{sec3} is devoted to the linear stability analysis around the HCS. This Section presents the main results of the paper. The dependence of the critical size $L_c$ on the parameter space is widely investigated in section \ref{sec4} by varying the parameters of the system in the case of a common coefficient of restitution ($\al_{11}=\al_{22}=\al_{12}\equiv \al$). The paper is closed in section \ref{sec5} with a brief discussion of the results.

\section{Hydrodynamic description}
\label{sec2}

We consider a binary mixture of \emph{inelastic}, smooth, hard spheres ($d=3$) or disks ($d=2$) of masses $m_{1}$ and $m_{2}$, and diameters $\sigma _{1}$ and $\sigma _{2}$.
The inelasticity of collisions among all pairs is characterized by three
independent constant coefficients of normal restitution $\alpha _{11}$,
$\alpha _{22}$, and $\alpha _{12}=\alpha _{21}$, where $\alpha _{ij}$ is the
coefficient of restitution for collisions between particles of species $i$
and $j$. At a kinetic level, all the relevant information on the state of the mixture is given through the one-particle velocity distribution function of each species $f_i(\mathbf{r},\mathbf{v},t)$ ($i=1,2$). This quantity gives the average number of particles of species $i$ that at instant $t$ are located around the point $\mathbf{r}$ with a velocity about $\mathbf{v}$. At moderate densities, the distribution functions $f_i$ are accurately described by
the coupled set of \emph{inelastic} Enskog kinetic equations \cite{GS95,BDS97}. From this set one can derive the (macroscopic) hydrodynamic equations for the particle number density of each species,
\beq
\label{2.0}
n_{i}\left( \mathbf{r},t\right)=\int\dd\mathbf{v} f_i(\mathbf{r},\mathbf{v},t),
\eeq
the mean flow velocity
\beq
\label{2.0.1}
\mathbf{U}\left( \mathbf{r},t\right)=\frac{1}{\rho\left( \mathbf{r},t\right)}
\sum_{i=1}^2\int\dd\mathbf{v} m_i \mathbf{v} f_i(\mathbf{r},\mathbf{v},t),
\eeq
and the \emph{granular} temperature
\beq
\label{2.0.2}
T\left(\mathbf{r},t\right)=\frac{2}{d n\left(\mathbf{r},t\right)}
\sum_{i=1}^2\int \dd\mathbf{v} m_i \left(\mathbf{v}-\mathbf{U}\left( \mathbf{r},t\right)\right)^2 f_i(\mathbf{r},\mathbf{v},t).
\eeq
The hydrodynamic equations are given by \cite{GDH07}:
\begin{equation}
D_{t}n_{i}+n_{i}\nabla \cdot \mathbf{U}+\frac{\nabla \cdot \mathbf{j}_{i}}{
m_{i}}=0, \quad i=1,2  \label{2.1}
\end{equation}
\begin{equation}
D_{t}\mathbf{U}+\rho ^{-1}\nabla \cdot \mathsf{P}=0\;,
\label{2.2}
\end{equation}
\begin{equation}
D_{t}T-\frac{T}{n}\sum_{i=1}^2\frac{\nabla \cdot \mathbf{j}_{i}}{m_{i}}+\frac{2}{
dn}\left( \nabla \cdot \mathbf{q}+\mathsf{P}:\nabla \mathbf{U}\right)
=-\zeta T\;.  \label{2.3}
\end{equation}
In the above equations, $D_{t}=\partial_{t}+\mathbf{U}\cdot
\nabla $ is the material derivative, $\rho =m_{1}n_{1}+m_{2}n_{2}$ is the total mass
density, $n=n_1+n_2$ is the total number density, $\mathbf{j}_{i}$ is the mass flux for species $i$, $\mathbf{q}$ is the heat flux, $\mathsf{P}$ is the pressure tensor, and
$\zeta $ is the cooling rate. In addition, the mass fluxes $\mathbf{j}_{1}$ and $\mathbf{j}_{2}$ are not independent since $\mathbf{j}_{2}=-\mathbf{j}_{1}$.

For the two component mixture considered here there are $d+3$ independent fields, $n_{1},$ $n_{2}$, $T$, and $\mathbf{U}$. To obtain a closed set of hydrodynamic equations, expressions for $\mathbf{j}_{i}$, $\mathbf{q}$, $\mathsf{P}$, and $\zeta$ must be given in terms of these fields. Such expressions are called ``constitutive equations''. These equations have been obtained up to the Navier--Stokes order from the Enskog equation in Ref.\ \cite{GDH07}. They are given by
\begin{equation}
\mathbf{j}_{1}=-\frac{m_1^2n_{1}}{\rho} D_{11}\nabla \ln n_{1}-
\frac{m_1 m_2n_{2}}{\rho} D_{12}\nabla \ln n_{2}
-\rho D^{T}\nabla \ln T,  \label{2.4}
\end{equation}
\begin{equation}
\label{2.5}
{\bf q}=-T^2D_{q,1}\nabla \ln n_1-T^2 D_{q,2}\nabla \ln n_2-\lambda
\nabla T,
\end{equation}
\begin{equation}
P_{k\ell }=p\delta _{k\ell}-\eta \left( \nabla_{\ell}U_{k}+\nabla
_{k}U_{\ell}-\frac{2}{d}\delta_{k\ell }\nabla \cdot \mathbf{U}\right)-\kappa
\delta_{k\ell }\nabla \cdot \mathbf{U},
\label{2.6}
\end{equation}
\begin{equation}  \label{2.7}
\zeta =\zeta^{(0)}+\zeta _{u}\nabla \cdot \mathbf{U}.
\end{equation}
Here, $D_{ij}$ are the mutual diffusion coefficients, $D^T$ is the thermal diffusion coefficient, $D_{q,ij}$ are the Dufour coefficients, $\lambda$ is the thermal conductivity coefficient, $p$ is the hydrostatic pressure, $\eta$ is the shear viscosity coefficient, $\kappa$ is the bulk viscosity coefficient, and $\zeta^{(0)}$ and $\zeta_u$ are the zeroth- and first-order contributions to the cooling rate, respectively. The eight transport coefficients $\left\{
D_{ij}, D^T, D_{q,ij}, \lambda, \eta, \kappa, \right\}$ as well as the first-order contribution $\zeta_u$ to the cooling rate verify a set of coupled linear integral equations which can be solved approximately by using the leading terms in a Sonine polynomial expansion. This type of solution provides explicit expressions for the complete set of Navier-Stokes transport coefficients and the cooling rate in terms of the volume fraction $\phi$, the coefficients of restitution $\alpha_{ij}$ and the parameters of the mixture (masses, sizes, and composition) \cite{GHD07,MGH12}.
The solid volume fraction $\phi$ is defined as $\phi=\phi_1+\phi_2$ where
\beq
\label{5.28.7}
\phi_i=\frac{\pi^{d/2}}{2^{d-1}d\Gamma\left(\frac{d}{2}\right)}n_i \sigma_i^d, \quad i=1,2.
\eeq

\begin{figure}
\includegraphics[width=0.75 \columnwidth,angle=0]{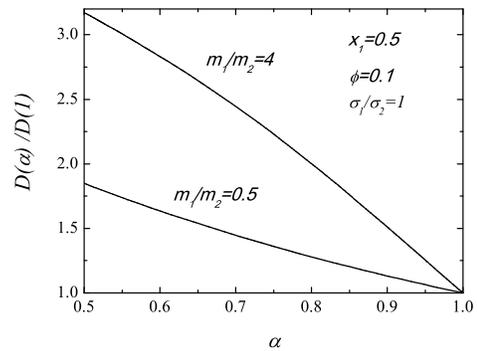}
\caption{Plot of the (reduced) diffusion coefficient $D(\al)/D(1)$ as a function of the (common) coefficient of restitution $\al$ for a binary mixture of inelastic hard spheres ($d=3$) with $\sigma_1=\sigma_2$, $x_1=0.5$, $\phi=0.1$ and two different values of the mass ratio $m_1/m_2$.
\label{fig1}}
\end{figure}
\begin{figure}
\includegraphics[width=0.75 \columnwidth,angle=0]{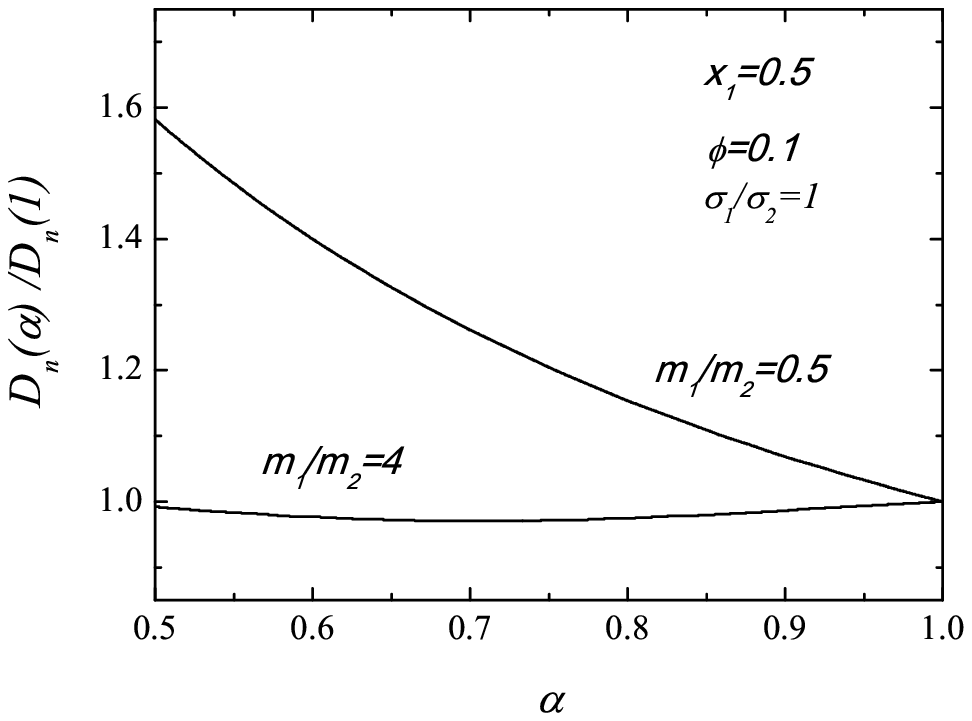}
\caption{Plot of the (reduced) density diffusion coefficient $D_n(\al)/D_n(1)$ as a function of the (common) coefficient of restitution $\al$ for a binary mixture of inelastic hard spheres ($d=3$) with $\sigma_1=\sigma_2$, $x_1=0.5$, $\phi=0.1$ and two different values of the mass ratio $m_1/m_2$.
\label{fig2}}
\end{figure}
\begin{figure}
\includegraphics[width=0.75 \columnwidth,angle=0]{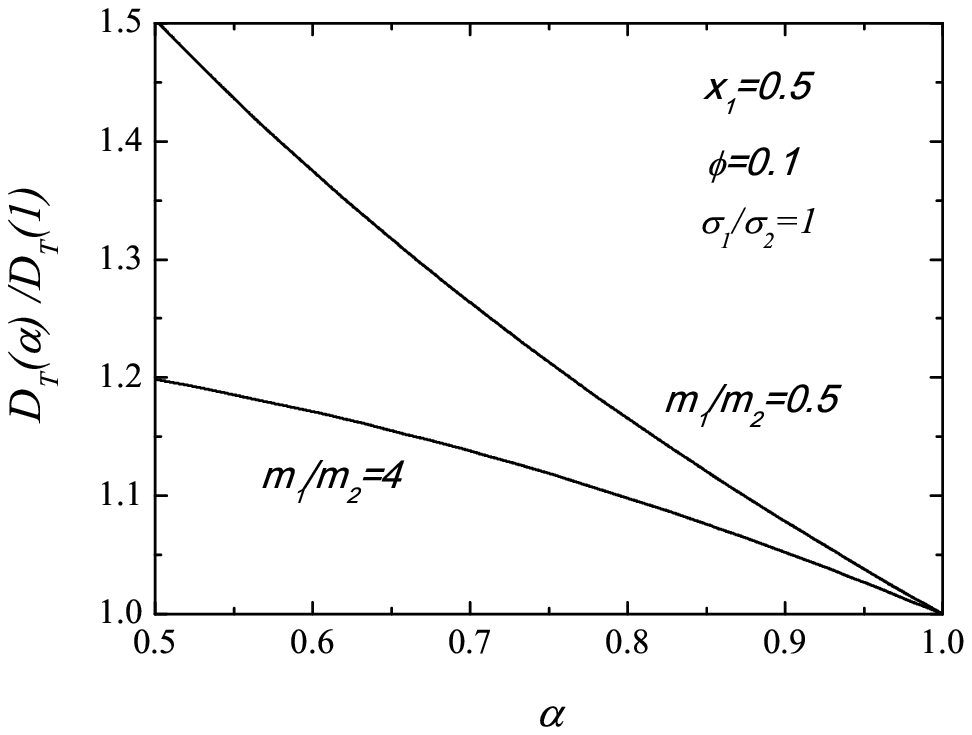}
\caption{Plot of the (reduced) thermal diffusion coefficient $D_T(\al)/D_{T}(1)$ as a function of the (common) coefficient of restitution $\al$ for a binary mixture of inelastic hard spheres ($d=3$) with $\sigma_1=\sigma_2$, $x_1=0.5$, $\phi=0.1$ and two different values of the mass ratio $m_1/m_2$.
\label{fig3}}
\end{figure}
\begin{figure}
\includegraphics[width=0.75 \columnwidth,angle=0]{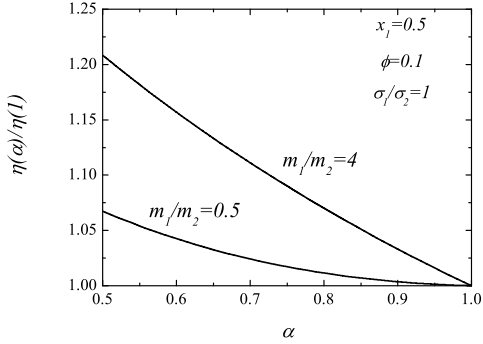}
\caption{Plot of the (reduced) shear viscosity coefficient $\eta(\al)/\eta(1)$ as a function of the (common) coefficient of restitution $\al$ for a binary mixture of inelastic hard spheres ($d=3$) with $\sigma_1=\sigma_2$, $x_1=0.5$, $\phi=0.1$ and two different values of the mass ratio $m_1/m_2$.
\label{fig4}}
\end{figure}

On the other hand, the expressions \eqref{2.4} and \eqref{2.5} for mass and heat fluxes can be defined in a variety of equivalent ways depending on the choice of the driving forces used. Here, to recover previous results \cite{G05} derived in the stability analysis for a monodisperse dense gas, the hydrodynamic fields $x_1$ and $n$ are considered instead of the partial densities $n_1$ and $n_2$. Here, $x_1=n_1/(n_1+n_2)$ is the composition (or mole fraction) of species 1. With this simple change of variables, Eqs.\ \eqref{2.1} for $n_1$ and $n_2$ become
\begin{equation}
\label{2.8}
D_t x_1+\frac{\rho}{n^2m_1m_2}\nabla \cdot \mathbf{j}_1=0,
\end{equation}
\begin{equation}
\label{2.9}
D_t n+n\nabla \cdot \mathbf{U}+\frac{m_2-m_1}{m_1m_2}\nabla \cdot \mathbf{j}_1=0.
\end{equation}
In terms of $\nabla x_1$ and $\nabla n$, the mass and heat fluxes read
\begin{equation}
\mathbf{j}_{1}=-\frac{m_1 m_{2}n}{\rho} D\nabla x_{1}-
\frac{m_1 m_2}{\rho} D_{n}\nabla n
-\frac{\rho}{T} D^{T}\nabla  T, \label{2.10}
\end{equation}
\begin{equation}
\label{2.11}
{\bf q}=-T^2D''\nabla  x_1-\frac{T^2}{n} D_{qn}\nabla  n-\lambda
\nabla T,
\end{equation}
where the transport coefficients $D$, $D_n$, $D"$, and $D_{qn}$ are defined as
\begin{equation}
\label{2.12}
D=\mu D_{11}-D_{12}, \quad D_n=x_1 \mu D_{11}+x_2 D_{12},
\end{equation}
\begin{equation}
\label{2.13}
D''=\frac{D_{q,1}}{x_1}-\frac{D_{q,2}}{x_2}, \quad D_{qn}=D_{q,1}+ D_{q,2},
\end{equation}
where $x_2=1-x_1$ and $\mu \equiv m_1/m_2$ is the mass ratio.

The explicit expressions of the Navier-Stokes transport coefficients of a $d$-dimensional dense granular binary mixture can be found in Ref.\ \cite{MGH12}. In the case of mechanically equivalent particles ($m_1=m_2$, $\sigma_1=\sigma_2$, and $\alpha_{ij}\equiv \alpha$), then $D_n=D^T=D''=0$ and Eqs.\ \eqref{2.6}, \eqref{2.10} and \eqref{2.11} agree with the results derived for a monodisperse dense gas \cite{GD99,L05}. Moreover, in the low-density limit ($\phi=0$), $\kappa=\zeta_u=0$ and the results derived for a granular binary mixture at low-density are recovered \cite{GD02,GM07}. Beyond the above two particular situations, the dependence of the transport coefficients on the parameter space of the system is quite intricate. To illustrate this dependence, Figs.\ \ref{fig1}--\ref{fig4} show the dimensionless quantities $D(\alpha)/D(1)$, $D_n(\alpha)/D_n(1)$, $D^T(\alpha)/D^T(1)$ and $\eta(\alpha)/\eta(1)$, respectively, as functions of the (common) coefficient of restitution $\alpha_{11}=\alpha_{22}=\alpha_{12}\equiv \alpha$ for $\phi=0.1$, $\sigma_1/\sigma_2=1$, $x_1=0.5$ and two different values of the mass ratio $m_1/m_2$. Here, $D(1)$, $D_n(1)$, $D^T(1)$ and $\eta(1)$ correspond to the values of these coefficients in the elastic limit. We observe that in general the influence of inelasticity on the transport coefficients is quite significant and so their functional form differs appreciably from their elastic form. This means that the previous predictions made for nearly elastic spheres \cite{JM89,Z95,AW98,AJ04} might quantitatively differ from those obtained here as the rate of dissipation increases. This will be confirmed later.

When the expressions \eqref{2.6}, \eqref{2.7}, \eqref{2.10}, and \eqref{2.11} for the pressure tensor, the cooling rate, the mass flux and the heat flux, respectively, are substituted into the \emph{exact} balance equations \eqref{2.2}, \eqref{2.3}, \eqref{2.8}, and \eqref{2.9} one gets the corresponding Navier-Stokes hydrodynamic equations for the hydrodynamic fields $x_1$, $n$, $\mathbf{U}$, and $T$. They are given by
\beqa
D_{t}x_{1}&=&\frac{\rho}{n^{2}m_{1}m_{2}}\nabla \cdot \left( \frac{
m_{1}m_{2}n}{\rho}D\nabla x_{1}+\frac{
m_{1}m_{2}}{\rho}D_{n}\nabla n\right. \nn
& &\left. +\frac{\rho }{T}D^T\nabla T\right),  \label{2.15}
\eeqa
\beqa
D_{t}n+n\nabla \cdot \mathbf{U}&=&\frac{m_2-m_1}{m_{1}m_{2}}\nabla \cdot
\left( \frac{m_{1}m_{2}n}{\rho}D\nabla x_{1}\right.\nn
& & \left.+\frac{m_{1}m_{2}}{\rho}D_{n}\nabla n+\frac{\rho }{T}D^T\nabla T\right),  \nn
\label{2.16}
\eeqa
\beqa
\rho D_{t}U_{\ell}+\nabla_{\ell}p &=&\nabla_{k}\left[\eta \left(
\nabla_{\ell }U_{k}+\nabla_{k}U_{\ell}-\frac{2}{d}\delta_{k\ell }\nabla
\cdot \mathbf{U}\right)\right.\nn
& &\left.
+\kappa \delta_{k\ell}\nabla \cdot \mathbf{U}\right],  \label{2.18}
\eeqa
\begin{eqnarray}
& &n\left(D_{t}+\zeta^{(0)}\right) T+\frac{2}{d}p\nabla \cdot \mathbf{U} =-\frac{T(m_2-m_1)}{m_1m_2}\nabla \nn
& & \cdot \left( \frac{
m_{1}m_{2}n}{\rho}D\nabla x_{1}+\frac{m_{1}m_{2}}{\rho}D_{n}\nabla n+
\frac{\rho }{T}D^T\nabla T\right)  \nonumber \\
&&+\frac{2}{d}\nabla \cdot \left( T^{2}D^{\prime \prime}\nabla
x_{1}+\frac{T^2}{n}D_{qn}\nabla n+\lambda \nabla T\right)  \nonumber \\
&&+\frac{2}{d} \left[ \eta\left(\nabla_{\ell }U_{k}+\nabla_{k}U_{\ell}-\frac{2
}{d}\delta_{k\ell}\nabla \cdot \mathbf{U}\right)+\delta_{k\ell}\kappa\nabla \cdot \mathbf{U}\right] \nn
& & \nabla_{\ell}U_{k}-nT\zeta_u\nabla \cdot \mathbf{U}.
\label{2.17}
\end{eqnarray}
Note that consistency would require to consider up to second order in the spatial gradients in the expression \eqref{2.7} for the cooling rate, since this is the order of the terms appearing in the energy balance equation \eqref{2.17} coming from the mass flux, the pressure tensor and the heat flux. Thus, since the cooling rate $\zeta$ is a scalar, its most general form at this order for a granular binary dense mixture is
\beqa
\label{burnett}
\zeta&=&\zeta^{(0)}+\zeta _{u}\nabla \cdot \mathbf{U}+\zeta_{n_1}\nabla^2 n_1+
\zeta_{n_2}\nabla^2 n_2+\zeta_T \nabla^2 T \nonumber\\
& & +\zeta_{TT}(\nabla T)^2+\zeta_{n_1 n_1}(\nabla n_1)^2+\zeta_{n_2 n_2}(\nabla n_2)^2 \nonumber\\
& & +\zeta_{T n_1}(\nabla T)\cdot (\nabla n_1)+\zeta_{T n_2}(\nabla T)\cdot (\nabla n_2)\nonumber\\
& &+\zeta_{n_1 n_2}(\nabla n_1)\cdot (\nabla n_2) 
+\zeta_{1,uu} (\nabla_i U_j)(\nabla_i U_j)\nonumber\\
& & +\zeta_{2,uu} (\nabla_i U_j)(\nabla_j U_i).
\eeqa
The first (linear) second-order terms ($\zeta_{n_i}$ and $\zeta_T$) have been determined for a one-component \emph{dilute} gas in Ref.\ \cite{BDKS98} while \emph{all} the set of coefficients (linear and nonlinear terms) have been computed for
granular monodisperse gases of viscoelastic particles in Ref.\ \cite{BP03}. The evaluation of the above set of coefficients for granular mixtures is a quite intricate problem. To the best of my knowledge, no explicit results for these coefficients have been reported for granular binary mixtures, even in the simplest case of a low-density mixture ($\phi=0$). On the other hand, it has
been shown for dilute gases that the contributions of the second-order terms to the cooling rate $\zeta$ are negligible \cite{BDKS98}, as compared with the corresponding zeroth-order contribution $\zeta^{(0)}$ (the first-order contribution $\zeta_u$ vanishes for dilute gases). It is assumed here that the same holds in the dense case
and so, for practical applications these second-order contributions can be in principle neglected in the Navier-Stokes hydrodynamic equations. In fact, the good agreement found in Ref.\ \cite{MGH14} between the present theoretical results (where the nonlinear contributions to $\zeta$ are not accounted for) and MD simulations for the onset of velocity vortices for strong inelasticity, finite density, and particle dissimilarity (see Figs.\ 2 and 3 of \cite{MGH14}) supports the above expectation.

The form of the Navier-Stokes hydrodynamic equations \eqref{2.15}--\eqref{2.17} is the same as for an ordinary binary mixture ($\alpha_{ij}=1$), except for the presence of the contributions to the cooling rate $\zeta_0$ and $\zeta_u$ and the dependence of the transport coefficients on the coefficients of restitution. This dependence is clearly illustrated in Figs.\ \ref{fig1}--\ref{fig4}.

\section{Stability of the linearized hydrodynamic equations}
\label{sec3}

In contrast to ordinary fluids, the hydrodynamic equations \eqref{2.15}--\eqref{2.17} admit nontrivial solutions even for spatially homogeneous states. This state is usually called homogeneous cooling state (HCS). In this case (no spatial gradients), Eqs.\ \eqref{2.15}--\eqref{2.17} read
\begin{equation}
\label{3.1}
\partial _{t}x_{1\text{H}}=\partial_t n_\text{H}=\partial _{t}u_{\text{H}\ell}=0,
\end{equation}
\begin{equation}
\left[ \partial_{t}+\zeta^{(0)} \left(x_{1\text{H}}, n_\text{H}, T_{\text{H}}\right) \right] T_{\text{H}}=0,
\label{3.2}
\end{equation}
where the subscript $\text{H}$ denotes the homogeneous state. The dependence of the zeroth-order cooling rate $\zeta^{(0)}=\zeta_1^{(0)}=\zeta_2^{(0)}$ on $x_{1\text{H}}$, $n_\text{H}$, and $T_{\text{H}}$ can be estimated by taking Maxellian distributions for the distributions $f_i$ in the HCS. Here, $\zeta_i^{(0)}$ is the partial cooling rate associated with the partial temperature $T_i$, which is a measure of the mean kinetic energy of particles of species $i$. The expression of $\zeta_i^{(0)}$ in the Maxwellian approximation is \cite{GD99}
\beqa
\label{3.3}
\zeta_i^{(0)}&=&\frac{d+2}{d}\nu \sum_{j=1}^2\; x_j\chi_{ij} \left(\frac{\sigma_{ij}}{\sigma_{12}}\right)^{d-1}\left(\frac{\theta_i+\theta_j}{\theta_i\theta_j}\right)^{1/2}
\nn
&\times& \left(1+\alpha_{ij}\right)
\left[1-\frac{\mu_{ji}}{2}\left(1+\alpha_{ij}
\right)\frac{\theta_i+\theta_j}{\theta_j}\right],
\eeqa
where $\chi_{ij}$ is the pair distribution function at contact, $\mu_{ij}=m_i/(m_i+m_j)$, $\theta_i=m_iT/\overline{m}T_i$, $\sigma_{ij}=(\sigma_i+\sigma_j)/2$, $\overline{m}\equiv (m_1+m_2)/2$, and
\begin{equation}
\label{3.4}
\nu=\frac{\pi^{(d-1)/2}}{\Gamma \left(\frac{d}{2}\right)}\frac{8}{d+2}n\sigma_{12}^{d-1}
\sqrt{\frac{T}{\overline{m}}}
\end{equation}
is an effective collision frequency chosen to recover previous results found for a dense one-component granular gas \cite{G05}. Note that for mechanically equivalent particles ($m_1=m_2$, $\sigma_1=\sigma_2$), the collision frequency $\nu$ is associated with the elastic shear viscosity of a dilute gas. Due to inelasticity in collisions, energy equipartition is broken and so, in general $T_1 \neq T_2$. The temperature ratio $\gamma\equiv T_1/T_2$ is determined from the condition $\zeta_1^{(0)}=\zeta_2^{(0)}$ \cite{GD99}.

In order to solve Eq.\ \eqref{3.2} it is convenient to change to a new time variable defined as
\begin{equation}
\label{3.5}
\tau=\frac{1}{2}\int_{0}^t\; \nu(T(t')) \dd t'.
\end{equation}
In this new time variable, the integration of Eq.\ \eqref{3.2} yields
\beq
\label{3.5.1}
T(t)=T(0)e^{-2\zeta_0^*\tau}
\eeq
where $\zeta_0^*=\zeta^{(0)}/\nu$. To find the relation between the ``internal'' time (related to the average number of collisions suffered per particle) and the ``external'' time $t$, one integrates the relation for $d\tau$ using $\nu\sim \sqrt{T}$ and gets the usual Haff's law \cite{H83}
\begin{equation}
\label{3.6}
T(t)=\frac{T(0)}{\left(1+\frac{1}{2}\zeta^{(0)}(0) t\right)^2},
\end{equation}
where $\zeta^{(0)}(0)$ is the cooling rate at the initial time. The partial temperatures $T_i$ also have the same time dependence \eqref{3.6} but each with a different value \cite{GD99}.

On the other hand, computer simulations \cite{GZ93,M93} have clearly shown that the HCS is unstable with respect to long enough wavelength perturbations. To analyze this problem it is convenient to carry on a (linear) stability analysis of the nonlinear hydrodynamic equations \eqref{2.15}--\eqref{2.17} with respect to the homogeneous state for small initial excitations. For ordinary fluid mixtures such perturbations decay in time according to the hydrodynamic modes of diffusion (shear, thermal, mass) and damped sound propagation. The perturbation analysis made here is for fixed coefficients of restitution different from unity in the long wavelength limit. It will be seen that the corresponding hydrodynamic modes for a granular mixture \cite{GMD07} differs from those obtained for normal mixtures. In addition, as has been widely explained in some previous papers \cite{BDKS98,G05,BR13,GMD07}, the linearization of Eqs.\ \eqref{2.15}--\eqref{2.17} about the homogeneous base state leads to a set of coupled partial differential equations with coefficients that depend on time since the HCS is cooling. This time dependence can be eliminated through convenient changes in the time and space variables and a scaling of the hydrodynamic fields.

Let $\delta y_{\beta}(\mathbf{r},t)=y_{\beta}(\mathbf{r},t)-y_{\text{H}\beta
}(t)$ denote the deviation of $\{x_{1}, n, \mathbf{U}, T\}$ from their values
in the HCS. If the initial spatial perturbation is sufficiently small, then
for some initial time interval these deviations will remain small and the
hydrodynamic equations \eqref{2.15}--\eqref{2.17} can be linearized with
respect to $\delta y_{\beta}(\mathbf{r},t)$. As said before, to eliminate the time dependence we introduce the time variable $\tau$ (defined in Eq.\ \eqref{3.5}) and the space variable
\begin{equation}
\boldsymbol{\ell}= \frac{1}{2}\frac{\nu_\text{H}(t)}{v_\text{H}(t)}\mathbf{r},  \label{3.7}
\end{equation}
where $v_{\text{H}}(t)=\sqrt{T_{\text{H}}(t)/\overline{m}}$. According to Eq.\  \eqref{3.5}, the dimensionless time scale $\tau$ is therefore an average number of collisions per particle in the time interval between 0 and $t$. The unit length $\nu_\text{H}(t)/v_\text{H}(t)$ introduced in Eq.\  \eqref{3.7} is proportional to the effective time-independent mean free path $1/n_\text{H} \sigma_{12}^{d-1}$.

A set of Fourier transformed dimensionless variables are then defined by
\begin{equation}
\rho_{1,\mathbf{k}}(\tau)=\frac{\delta x_{1\mathbf{k}}(\tau)}{x_{1\text{H}}},\quad
\rho_{\mathbf{k}}(\tau)=\frac{\delta n_{\mathbf{k}}(\tau)}{n_{\text{H}}},  \label{3.9}
\end{equation}
\begin{equation}
\mathbf{w}_{\mathbf{k}}(\tau)=\frac{\delta \mathbf{U}_{\mathbf{k}}(\tau)}{
v_{\text{H}}(\tau)}, \quad
\theta_{\mathbf{k}}(\tau)=\frac{\delta T_{\mathbf{k}}(\tau)}{T_{\text{H}}(\tau)},
\label{5.5}
\end{equation}
where $\delta y_{\mathbf{k}\beta}(\tau)\equiv \left\{\rho_{1,\mathbf{k}}(\tau), \rho_{\mathbf{k}}(\tau), \mathbf{w}_{\mathbf{k}}(\tau), \theta_{\mathbf{k}}(\tau)\right\}$ is defined as
\begin{equation}
\delta y_{\mathbf{k}\beta}(\tau)=\int \dd\boldsymbol{\ell}\;e^{-i\mathbf{k} \cdot \boldsymbol{\ell}}
\delta y_{\beta}(\boldsymbol{\ell},\tau).
\label{3.8}
\end{equation}
Note that in Eq.\ \eqref{3.8} the wave vector $\mathbf{k}$ is dimensionless. In terms of these variables, the $d-1$ transverse velocity components  ${\bf w}_{{\bf k}\perp}={\bf w}_{{\bf k}}-({\bf w}_{{\bf k}}\cdot
\widehat{{\bf k}})\widehat{{\bf k}}$ (orthogonal to the wave vector ${\bf k}$)
decouple from the other four modes and hence can be obtained more
easily. Their evolution equation is
\begin{equation}
\label{5.6}
\left(\frac{\partial}{\partial \tau}-\zeta_0^*+\frac{1}{2}\eta^*
k^2\right){\bf w}_{{\bf k}\perp}=0,
\end{equation}
where
\begin{equation}
\label{5.7}
\eta^*\equiv \frac{\nu_\text{H} \eta_\text{H}}{\rho_\text{H} v_\text{H}^2}.
\end{equation}
In Eq.\ \eqref{5.6} it is understood that $\zeta_0^*$ is also evaluated in the HCS.
The solution to Eq.\ (\ref{5.6}) is
\begin{equation}
\label{5.8}
{\bf w}_{{\bf k}\perp}({\bf k}, \tau)={\bf w}_{{\bf k}\perp}(0)\exp[s_{\perp}(k)\tau],
\end{equation}
where
\begin{equation}
\label{5.9}
s_{\perp}(k)=\zeta_0^*-\frac{1}{2}\eta^* k^2.
\end{equation}
This identifies $d-1$ shear (transversal) modes analogous to the
elastic ones \cite{RL}. According to Eq.\ (\ref{5.9}), there exists
a critical wave number $k_{\perp}^c$ given by
\begin{equation}
\label{5.10}
k_{\perp}^c=\left(\frac{2\zeta_0^*}{\eta^*}\right)^{1/2}.
\end{equation}
This critical value separates two regimes: shear modes with $k> k_{\perp}^c$ always decay
while those with $k< k_{\perp}^c$ grow exponentially.

The remaining (longitudinal) modes correspond to the composition field $\rho_{1,\mathbf{k}}$, the density field $\rho_{\mathbf{k}}$, the longitudinal component of the velocity field $\mathbf{w}_{\mathbf{k}||}={\bf w}_{{\bf k}}\cdot
\widehat{{\bf k}}$ (parallel to $\mathbf{k}$), and the temperature field $\theta_{\mathbf{k}}$. These modes are coupled and obey the time-dependent equation
\begin{equation}
\frac{\partial \delta z_{\mathbf{k}\beta}(\tau)}{\partial \tau}=\left(
M_{\beta\gamma}^{(0)}+ikM_{\beta\gamma}^{(1)}+k^{2}M_{\beta\gamma
}^{(2)}\right) \delta z_{\mathbf{k}\gamma}(\tau),  \label{5.24}
\end{equation}
where now $\delta z_{\mathbf{k}\beta}(\tau)$ denotes the four variables $
\left( \rho_{1,\mathbf{k}},\rho_{\mathbf{k}}, \theta_{\mathbf{k}}, w_{\mathbf{
k}||}\right)$. The matrices in Eq.\ \eqref{5.24} are
\begin{equation}
M^{(0)}=\left(
\begin{array}{cccc}
0 & 0 & 0 & 0 \\
0 &0&0&0\\
-2x_{1}\frac{\partial \zeta_0^*}{\partial x_{1}}&
-2\frac{\partial (n^* \zeta_0^*)}{\partial n^*}& -\zeta_0^{\ast}& 0\\
0 & 0 & 0 &\zeta_0^*
\end{array}
\right) ,  \label{5.25}
\end{equation}
\begin{widetext}
\begin{equation}
M^{(1)}=\left(
\begin{array}{cccc}
0 & 0 & 0 & 0 \\
0 & 0 & 0 & -1\\
0 & 0 & 0 & -\frac{2}{d}p^*-\zeta_u\\
-\frac{(1+\mu)\delta}{2(1+\mu\delta)}\frac{\partial p^*}{\partial x_{1}}
&-\frac{(1+\mu)\delta}{2x_2(1+\mu\delta)}\frac{\partial (n^*p^*)}{\partial n^*}&-\frac{(1+\mu)\delta}{2x_2(1+\mu\delta)}p^* & 0
\end{array}
\right) ,  \label{5.26}
\end{equation}
\begin{equation}
M^{(2)}=\left(
\begin{array}{cccc}
-\frac{x_2(1+\mu \delta)}{4\mu_{12}}D^{\ast} &-\frac{x_2(1+\mu \delta)}{4\mu_{12}}D_n^{\ast}
 &-\frac{x_2(1+\mu \delta)}{4\mu_{12}}D^{T\ast}& 0 \\
-\frac{x_1(1-\mu)}{4\mu_{12}}D^*&-\frac{x_1(1-\mu)}{4\mu_{12}}D_n^*&-\frac{x_1(1-\mu)}{4\mu_{12}}D^{T*}& 0\\
\frac{x_1(1-\mu)}{4\mu_{12}}D^*-\frac{d+2}{4}x_1D^{''*} & \frac{x_1(1-\mu)}{4\mu_{12}}D_n^*
-\frac{d+2}{4}D_{qn}^*& \frac{x_1(1-\mu)}{4\mu_{12}}D^{T*}-\frac{d+2}{4}\lambda^*& 0 \\
0 & 0 & 0 & -\frac{d-1}{d}\eta^*-\frac{1}{2}\kappa^*
\end{array}
\right),  \label{5.27}
\end{equation}
\end{widetext}
where the subscript $\text{H}$ has been omitted in Eqs.\ \eqref{5.25}--\eqref{5.27} for the sake of brevity. In these equations, $n^*\equiv n\sigma_{12}^d$, $\delta\equiv x_1/x_2$, and I have introduced the dimensionless transport coefficients
\begin{equation}
\label{5.16}
D^*\equiv \frac{m_1m_2\nu_\text{H}}{\rho_\text{H} T_\text{H}}D_\text{H}, \quad D_n^*\equiv \frac{m_1m_2\nu_\text{H}}{\rho_\text{H} T_\text{H}}D_{n,\text{H}},
\end{equation}
\beq
\label{5.16.1}
D^{T*}\equiv \frac{\rho_\text{H} \nu_\text{H}}{n_\text{H} T_\text{H}}D_\text{H}^T, \quad
D^{''*}\equiv \frac{4}{d(d+2)}\frac{\overline{m}\nu_\text{H}}{n_\text{H}}D_\text{H}'',
\eeq
\begin{equation}
\label{5.23}
D_{qn}^*\equiv \frac{4}{d(d+2)}\frac{\overline{m}\nu_\text{H}}{n_\text{H}}D_{qn,\text{H}},
\quad \lambda^{*}\equiv \frac{4}{d(d+2)}\frac{\overline{m}\nu_\text{H}}{n_\text{H} T_\text{H}}\lambda_\text{H},
\end{equation}
\begin{equation}
\kappa^{\ast}=\frac{\nu_{\text{H}}}{\rho_{\text{H}}v_{\text{H}}^{2}}\kappa_\text{H}. \label{5.20bis}
\end{equation}
In addition, the (reduced) hydrostatic pressure $p^*\equiv p/(nT)$ is given by \cite{GDH07}
\beq
\label{5.18}
p^*=1+\frac{\pi^{d/2}}{d\Gamma \left(\frac{d}{2}\right)}n^* \sum_{i,j}
x_i x_j  \left(\frac{\sigma_{j}}{\sigma_{12}}\right)^d \mu_{ji} \left(1+\alpha_{ij}\right)\chi_{ij}\gamma_i,
\eeq
where the temperature ratios $\gamma_i\equiv T_i/T$ are defined as
\beq
\label{5.18.1}
\gamma_1=\frac{\gamma}{1+x_1(\gamma-1)}, \quad
\gamma_2=\frac{1}{1+x_1(\gamma-1)}.
\eeq
Note that $p^*$ depends explicitly on $n^*$ and $x_1$ and also through its dependence on $\chi_{ij}$ and $\gamma_i$.

\begin{figure*}
%[hbtp]
\begin{center}
\begin{tabular}{lr}
\resizebox{7cm}{!}{\includegraphics{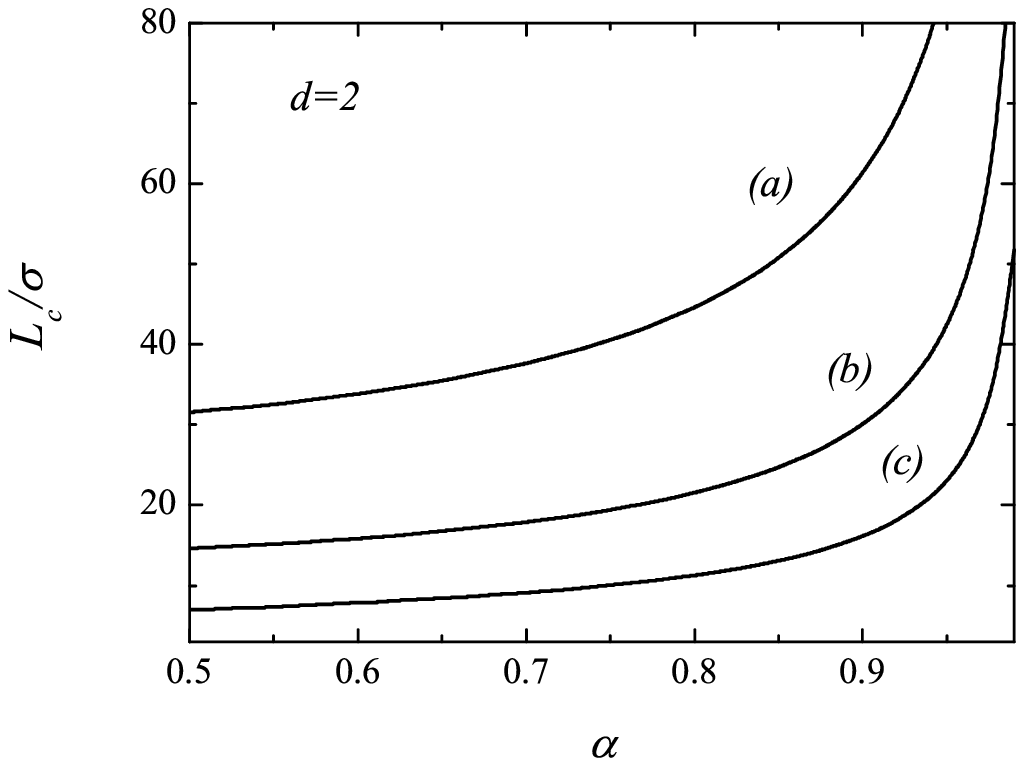}}&\resizebox{6.5cm}{!}
{\includegraphics{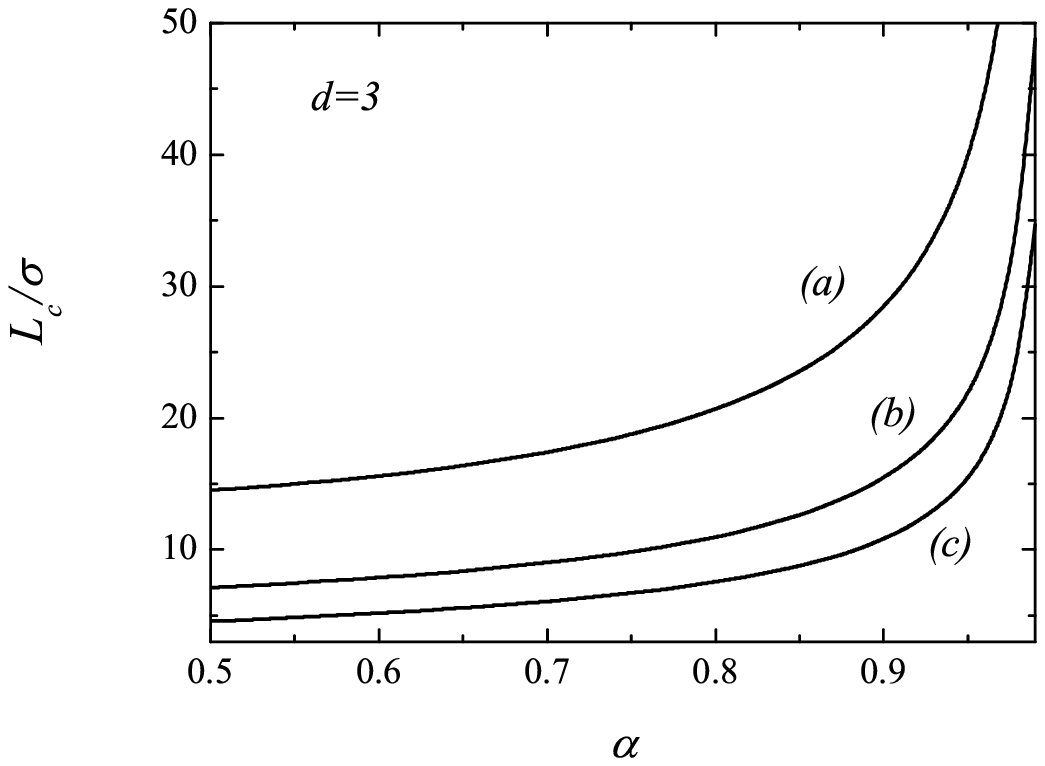}}
\end{tabular}
%\resizebox{6.5cm}{!}{\includegraphics{fig4.eps}}
\end{center}
\caption{The critical length scale $L_c$ for velocity vortices in units of the diameter $\sigma$ as a function of the coefficient of restitution $\al$ for a system of mechanically equivalent particles. Three different values of the volume fraction $\phi$ are considered: (a) $\phi=0.1$, (b) $\phi=0.2$, and (c) $\phi=0.4$. The left panel corresponds to disks ($d=2$) while the right panel refers to spheres ($d=3$). In each case, the system is linearly stable for points below the corresponding curve
\label{fig5}}
\end{figure*}

The longitudinal four modes have the form $\exp[s_{\text{n}}(k)\tau]$ for $n=1, 2, 3, 4$, where $s_{\text{n}}(k)$ are the eigenvalues of the matrix
\begin{equation}
\label{5.28.1}
{\sf M}\equiv {\sf M}^{(0)}+ik{\sf M}^{(1)}+k^2 {\sf M}^{(2)}.
\end{equation}
In other words, they are the solutions of the quartic equation $A(k,s)=0$ where
\begin{equation}
\label{5.28.2}
A(k,s)\equiv \det({\sf M}-s \openone).
\end{equation}
Here, $\openone$ denotes the identity matrix. Although the explicit form of $A(k,s)$ can be easily obtained from Eqs.\ \eqref{5.25}--\eqref{5.27}, its expression will be omitted here for the sake of brevity.

It is instructive to consider first the solutions to Eqs.\ \eqref{5.9} and \eqref{5.28.2} in the extreme long wavelength limit, $k=0$. In this case, the eigenvalues of the hydrodynamic modes are given by
\begin{equation}
s_{\perp}=\frac{1}{2}\zeta_0^{\ast},\quad s_{\text{n}}=\left(0, 0,
-\zeta_0^{\ast}, \zeta_0^{\ast}\right).
\label{5.28.3}
\end{equation}
Two of the eigenvalues are
positive, corresponding to growth of the initial perturbation in time. Thus,
some of the solutions are unstable. The two zero eigenvalues represent
marginal stability solutions, while the negative eigenvalue gives stable
solutions. For general initial perturbations all modes are excited. These
modes correspond to evolution of the fluid due to uniform perturbations of
the HCS, i.e. a global change in the HCS parameters. The unstable modes are
seen to arise from the initial perturbations $w_{\mathbf{k}\perp }(0)$ or
$w_{\mathbf{k}||}(0)$.  The unstable modes may appear trivial
since they are due entirely to the normalization of the fluid velocity by
the time dependent thermal velocity $v_H(t)$. However, this normalization is required
by the scaling of the entire set of equations to obtain time independent
coefficients.

For $k\neq 0$, the dependence of the longitudinal modes on the wave vector $k$ is quite intricate. On the other hand, the critical longitudinal mode $k_{||}^c$ can be obtained from the equation $A(k,s)=0$ when $s=0$. This leads to the quartic equation
\begin{equation}
\label{5.28.4}
A_0+A_2 k^2+A_4 k^4=0,
\end{equation}
where the coefficients $A_i$ are known functions of the solid volume fraction, the coefficients of restitution, and the parameters of the mixture. The critical value $k_{||}^c$ is the largest real root of Eq.\ \eqref{5.28.4}. As before, given that its explicit expression is very long and not relevant for the purposes of this paper, I shall omit the form of $k_{||}^c$ for the sake of simplicity.

For mechanically equivalent particles ($m_1=m_2\equiv m$, $\sigma_1=\sigma_2\equiv \sigma$, and $\alpha_{ij}\equiv \alpha$), the results derived in this section agree with those obtained before for \emph{monodisperse} dense gases \cite{G05}. Moreover, in the low-density limit ($n^*\to 0$), one recovers the dispersion relations obtained in a previous work \cite{GMD07} for a dilute granular binary mixture. These limits show the consistency of the linear stability analysis carried out in this paper for dense granular binary mixtures.

\section{Critical size for the onset of instabilities}
\label{sec4}

In a system with periodic boundary conditions, the smallest allowed wave
number is $2\pi/L$, where $L$ is the largest system length.
Hence, for given values of density, coefficients of restitution and parameters of the mixture, we can identify
a critical size $L_c$ so that the system becomes unstable when $L>L_c$. The value of $L_c$ is determined by equating
\begin{equation}
\label{5.28.5}
\frac{2\pi}{L_c^*}=\text{max}\{k_{\perp}^c,k_{||}^c\}, \quad L_c^*=\frac{\nu_\text{H}}{2 v_\text{H}}L_c.
\end{equation}
On the other hand, the present results show that in general $k_{\perp}^c>k_{||}^c$ and hence, the origin of instability is associated with the transversal components of the velocity field. This result has been verified for a low-density monodisperse granular gas \cite{BRM98} and for a dilute granular binary mixture \cite{BR13} by numerically solving the Boltzmann equation by means of the Direct Simulation Monte Carlo (DSMC) method \cite{B94}. Recent MD simulations \cite{MDCPH11,MGHEH12,MGH14} for granular fluids at moderate densities have also found that $k_{\perp}^c>k_{||}^c$. In this latter case, the theoretical predictions for the critical size compare well with MD simulations even for strong dissipation \cite{MGH14}. Thus, according to Eqs.\ \eqref{5.10} and \eqref{5.28.5}, the critical length scale for velocity vortex instability is given by
\beq
\label{5.28.6}
\frac{L_c}{\sigma_{12}}=\frac{d+2}{2\sqrt{2}}\frac{\Gamma\left(\frac{d}{2}\right)}{\pi^{\frac{d-3}{2}}}
\sqrt{\frac{\eta^*}{\zeta_0^*}}
\left(n_\text{H}\sigma_{12}^{d}\right)^{-1}.
\eeq
\begin{figure*}
%[hbtp]
\begin{center}
\begin{tabular}{lr}
\resizebox{7cm}{!}{\includegraphics{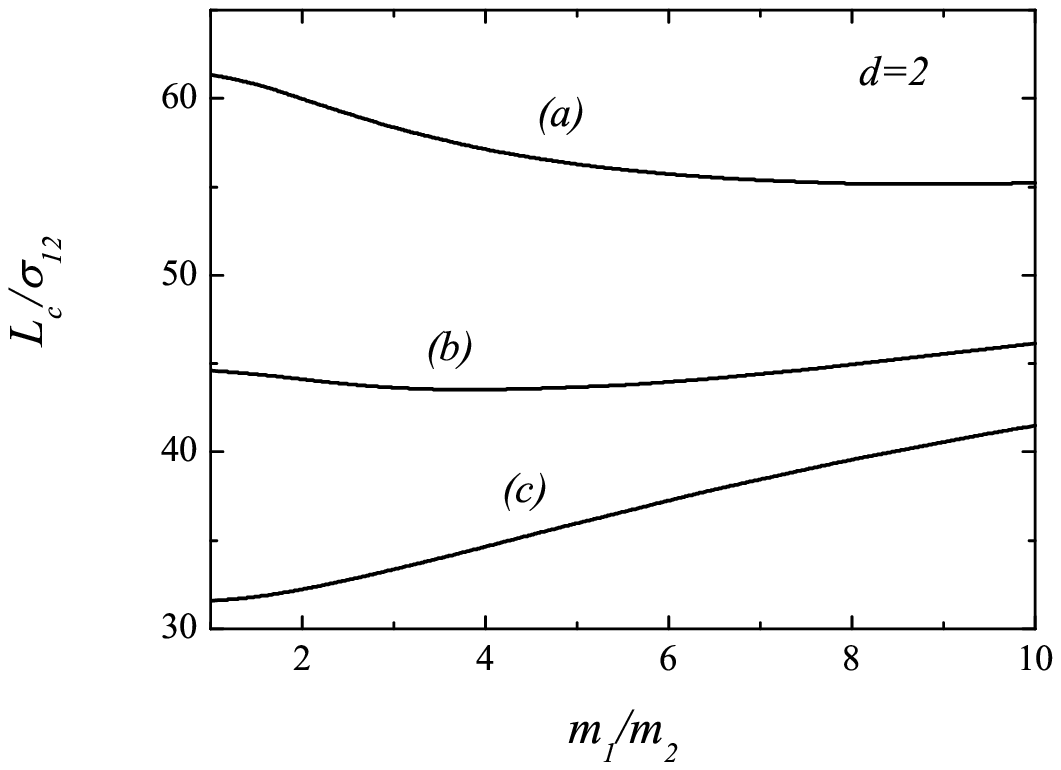}}&\resizebox{6.5cm}{!}
{\includegraphics{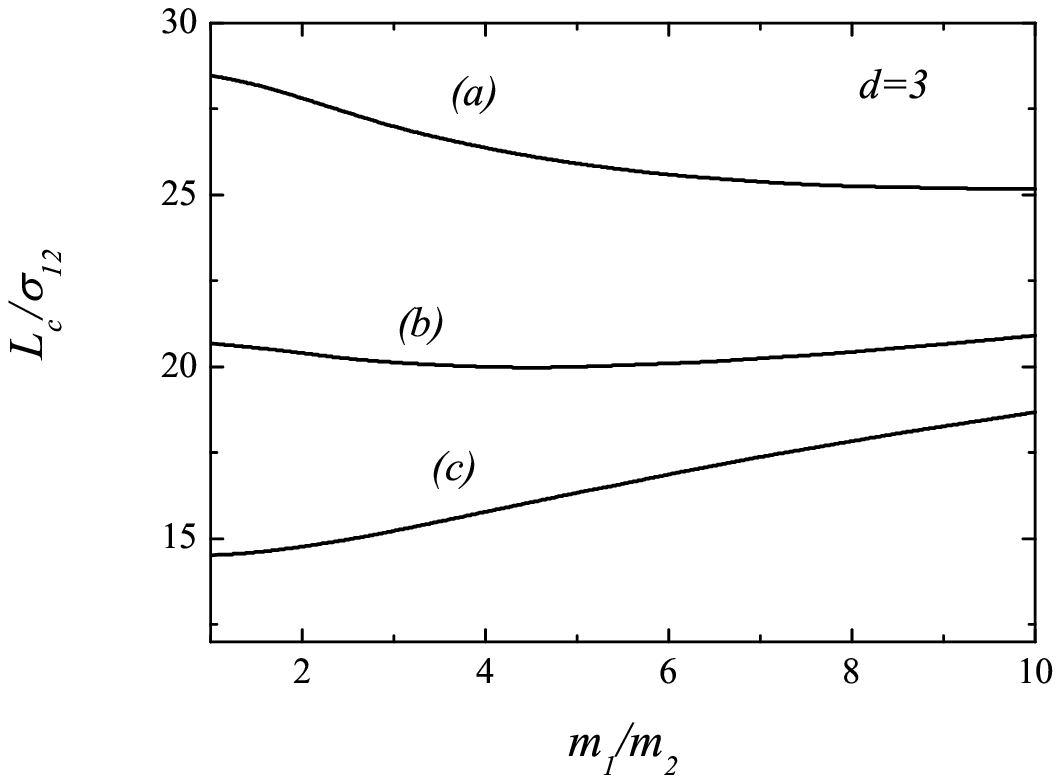}}
\end{tabular}
%\resizebox{6.5cm}{!}{\includegraphics{fig4.eps}}
\end{center}
\caption{The critical length scale $L_c$ for velocity vortices in units of $\sigma_{12}$ as a function of the mass ratio $m_1/m_2$. We have considered a granular binary mixture with $x_1=0.5$, $\sigma_1/\sigma_2=1$ and $\phi=0.1$. Three different values of the (common) coefficient of restitution are studied: (a) $\al=0.9$, (b) $\al=0.8$, and (c) $\al=0.5$. The left panel corresponds to disks ($d=2$) while the right panel refers to spheres ($d=3$). In each case, the system is linearly stable for points below the corresponding curve.
\label{fig6}}
\end{figure*}
\begin{figure*}
%[hbtp]
\begin{center}
\begin{tabular}{lr}
\resizebox{6.8cm}{!}{\includegraphics{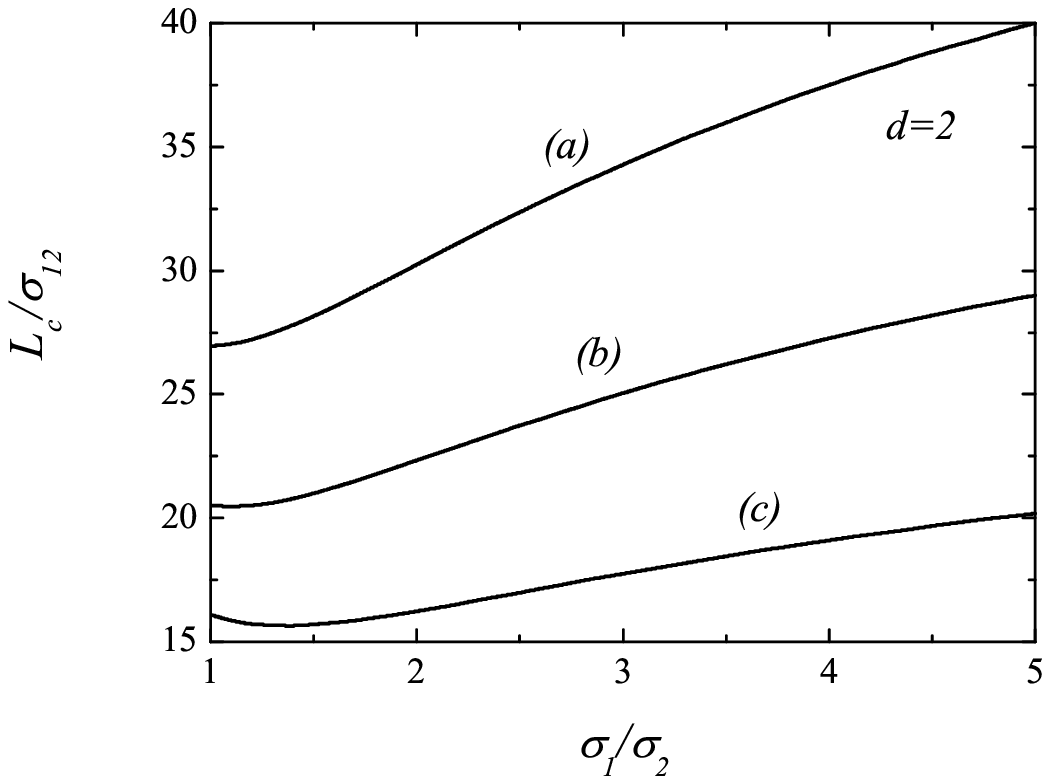}}&\resizebox{7cm}{!}
{\includegraphics{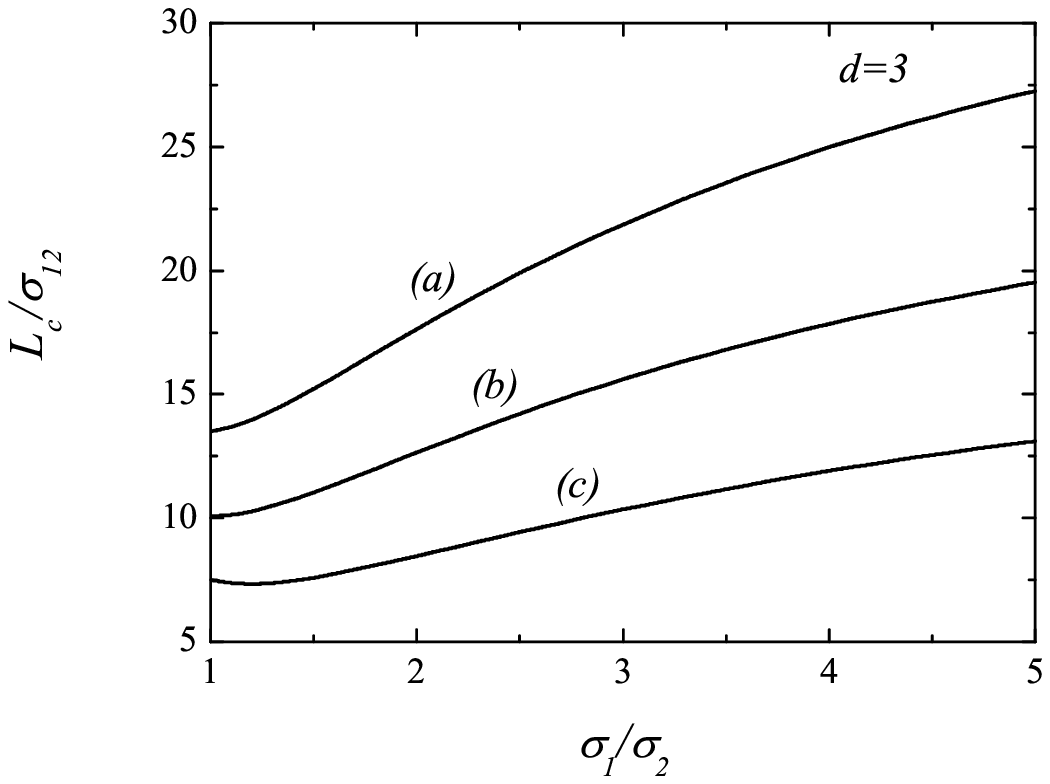}}
\end{tabular}
%\resizebox{6.5cm}{!}{\includegraphics{fig4.eps}}
\end{center}
\caption{The critical length scale $L_c$ for velocity vortices in units of $\sigma_{12}$ as a function of the ratio of diameters $\sigma_1/\sigma_2$. We have considered a granular binary mixture with $x_1=0.5$, $m_1/m_2=5$ and $\phi=0.2$. Three different values of the (common) coefficient of restitution are studied: (a) $\al=0.9$, (b) $\al=0.8$, and (c) $\al=0.5$. The left panel corresponds to disks ($d=2$) while the right panel refers to spheres ($d=3$). In each case, the system is linearly stable for points below the corresponding curve.
\label{fig7}}
\end{figure*}
\begin{figure*}
%[hbtp]
\begin{center}
\begin{tabular}{lr}
\resizebox{6.8cm}{!}{\includegraphics{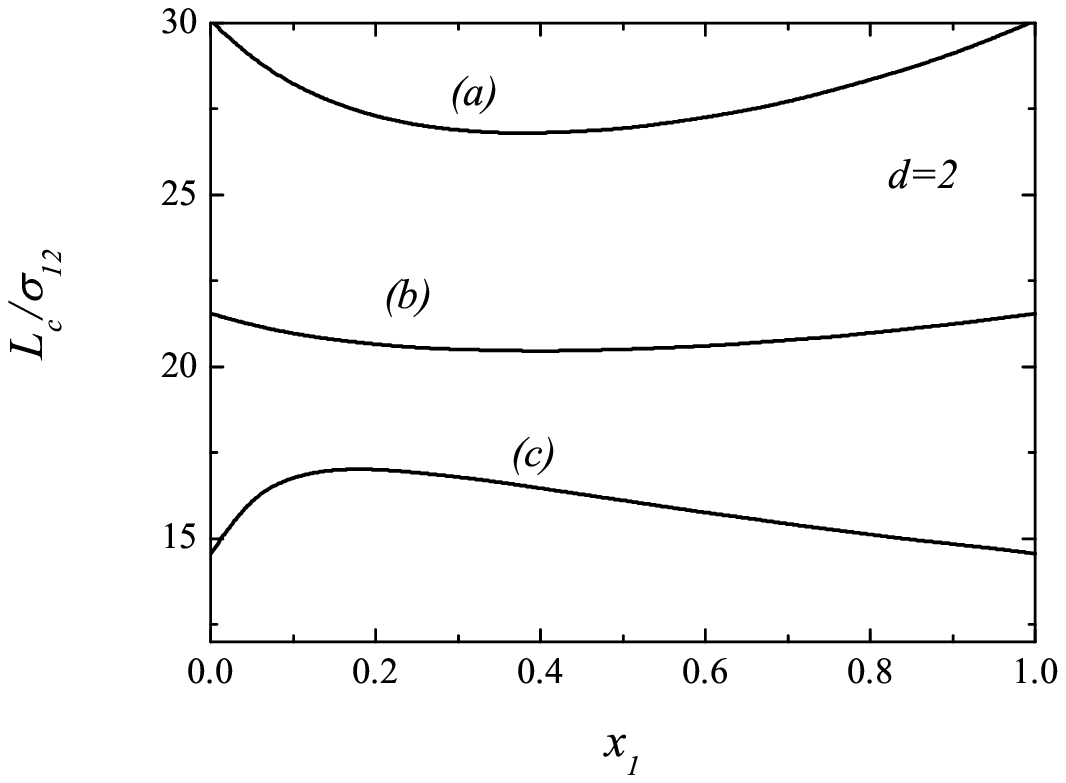}}&\resizebox{7cm}{!}
{\includegraphics{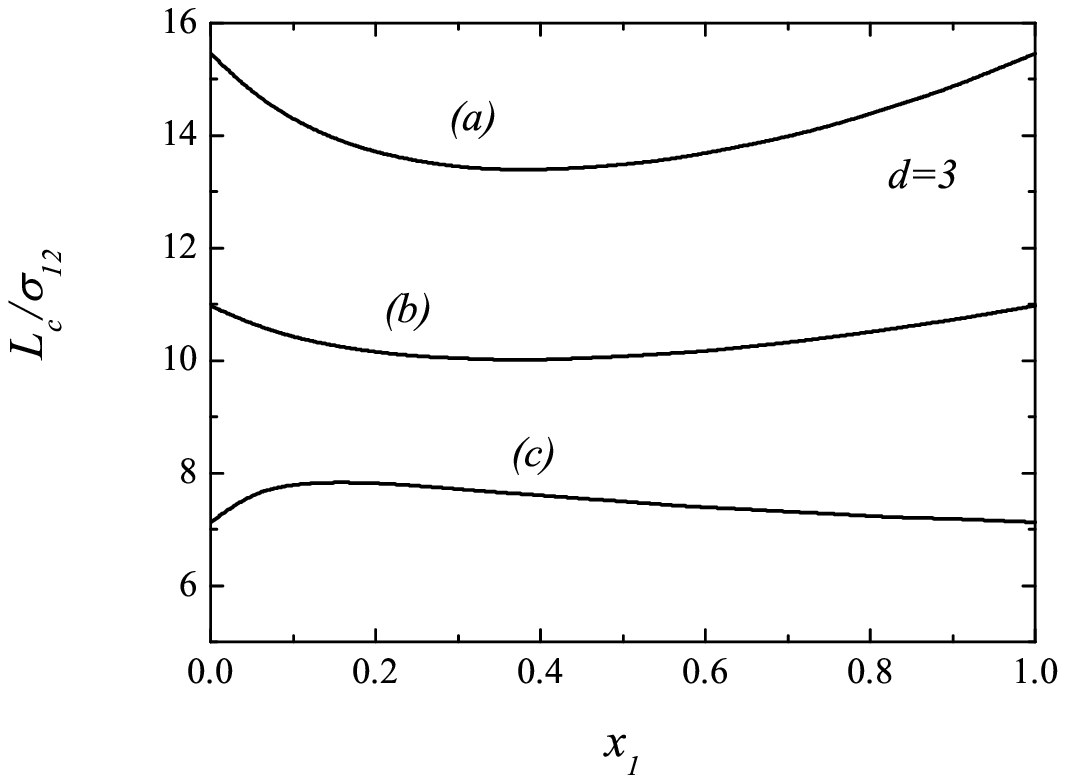}}
\end{tabular}
%\resizebox{6.5cm}{!}{\includegraphics{fig4.eps}}
\end{center}
\caption{The critical length scale $L_c$ for velocity vortices in units of $\sigma_{12}$ as a function of the concentration $x_1$. We have considered a granular binary mixture with $m_1/m_2=5$, $\sigma_1/\sigma_2=1$ and $\phi=0.2$. Three different values of the (common) coefficient of restitution are studied: (a) $\al=0.9$, (b) $\al=0.8$, and (c) $\al=0.5$. The left panel corresponds to disks ($d=2$) while the right panel refers to spheres ($d=3$). In each case, the system is linearly stable for points below the corresponding curve.
\label{fig8}}
\end{figure*}

Equation \eqref{5.28.6} gives the dependence of the critical length on the parameter space of the system. Specifically, the parameter space is the mass ratio $m_1/m_2$, the composition $n_1/(n_1+n_2)$, the ratio of diameters $\sigma_1/\sigma_2$, the coefficients of restitution $\al_{ij}$, and the solid volume fraction $\phi$.
According to Eq.\ \eqref{5.28.6}, the dependence of $L_c$ on the parameters of the mixture is essentially accounted for by the term $\sqrt{\eta^*/\zeta_0^*}$. The expression of the cooling rate $\zeta_0^*$ is given by Eq.\ \eqref{3.3} while the explicit form of the (reduced) shear viscosity $\eta^*$ is provided in the Appendix \ref{appA} for the sake of completeness.

Given that the parameter space of the problem is large, in order to reduce the number of independent parameters the simplest case of a \emph{common} coefficient of restitution ($\al\equiv \al_{11}=\al_{22}=\al_{12}$) is considered. Thus, once the dimensionality of the system is fixed, the parameter space is reduced to five dimensionless quantities: $\left\{m_1/m_2, \sigma_1/\sigma_2, x_1, \phi, \al \right\}$.

Three different values of the overall volume fractions have been considered: $\phi=0.1$, $\phi=0.2$, and $\phi=0.4$. The first two values of $\phi$ represent a granular fluid with moderate density while the latter one corresponds to a system with high density. Three different values of the common coefficient of restitution have been analyzed: $\al=0.9$ (weak dissipation), $\al=0.8$ (moderate dissipation), and $\al=0.5$ (strong dissipation).

Before considering a binary mixture, the case of mechanically equivalent particles (monodisperse dense gas) is illustrated for inelastic hard disks ($d=2$) and spheres ($d=3$). Figure \ref{fig5} shows $L_c/\sigma$ versus the coefficient of restitution $\al$ for different values of the volume fraction $\phi$. First, we observe that although the dependence of the critical length on dissipation is quite similar in $d=2$ and $d=3$, the magnitude of $L_c$ (measured in units of the diameter $\sigma$) is larger in disks than spheres. Thus, for a given value of $\al$, the critical length for velocity vortices increases as the dimensionality of the system decreases. With respect to the dependence on the density $\phi$, at a given value of dissipation, it is quite apparent that $L_c$ decreases with density and hence, smaller systems are required to observe the shearing instability as the granular mixture becomes denser.
\begin{figure}
\includegraphics[width=0.8 \columnwidth,angle=0]{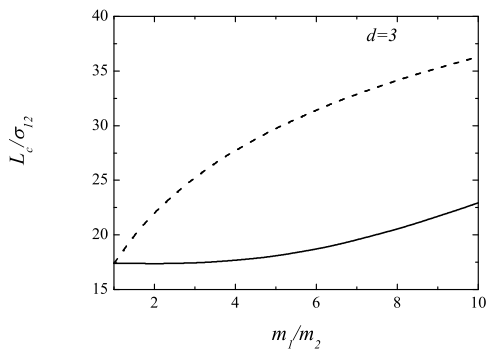}
\caption{The (dimensionless) critical length scale $L_c/\sigma_{12}$ as a function of the mass ratio $m_1/m_2$ for a binary mixture of inelastic hard spheres ($d=3$) with $x_1=0.1$, $\sigma_1/\sigma_2=1$, $\phi=0.1$ and a (common) coefficient of restitution $\al=0.7$. The solid line is the result derived here while the dashed line is the result derived from Eq.\ \eqref{5.28.6} by assuming energy equipartition ($\gamma=1$) and by neglecting the $\al$-dependence of $\eta^*$.
\label{fig9}}
\end{figure}

Now we consider granular binary mixtures. Figure \ref{fig6} shows $L_c/\sigma_{12}$ as a function of the mass ratio $m_1/m_2$ with $x_1=0.5$, $\sigma_1/\sigma_2=1$, $\phi=0.1$ and different values of $\al$. Regarding the influence of dimensionality on the critical length, for given values of the coefficient of restitution and the parameters of the mixture, the critical length is (significantly) larger for $d=2$ than for $d=3$. As will show later, this trend is also observed in the remaining plots presented in this paper. Moreover, at a given value of $m_1/m_2$, as expected $L_c$ decreases with collisional dissipation. In fact, $L_c\to \infty$ for elastic collisions ($\al=1$). The theoretical results also show that the $L_c/\sigma_{12}$ predictions for $\al=0.8$ and $\al=0.5$ seem to converge for large mass ratios and eventually crossover. This tendency is consistent with MD simulations (see for instance, fig.\ 2 of Ref.\ \cite{MGH14}). Figure \ref{fig7} shows the critical size as a function of the ratio of diameters with $x_1=0.5$, $m_1/m_2=5$ and $\phi=0.2$. It is quite apparent that $L_c/\sigma_{12}$ increases with the ratio $\sigma_1/\sigma_2$, except in a small region of $\sigma_1/\sigma_2$ close to one. In addition, the influence of the size ratio on the critical length is stronger for weak dissipation ($\al=0.9$) than for strong dissipation ($\al=0.5$), this behavior being independent of the dimensionality of system. Finally, the dependence of the critical length on species composition $x_1$ is illustrated in Fig. \ \eqref{fig8} for a binary mixture with $\sigma_1/\sigma_2=1$, $m_1/m_2=5$ and $\phi=0.2$. We see that $L_c$ is not a monotonic function of $x_1$. Moreover, the impact of composition on the critical size is much less important than the one observed with respect to the mass ratio (see Fig.\ \ref{fig6}) and/or the size ratio (see Fig.\ \ref{fig7}). In fact, for $\al=0.8$, $L_c$ displays a very weak dependence on $x_1$. It is important to recall that the theoretical predictions for $L_c$ derived from the (inelastic) Enskog kinetic equation have been recently assessed against MD simulations of inelastic hard spheres \cite{MGH14}. The comparison carried out in this work has shown in general an excellent agreement between theory and simulation when physical properties of particles are not quite disparate. This agreement becomes only good (at worst 20\% error) for the most extreme conditions studied [see for instance, Fig.\ 2(b) for $m_1/m_2=10$, $x_1=0.1$, $\phi=0.2$ and $\al=0.7$]. On the other hand, based on the results obtained for the tracer diffusion coefficient \cite{GV09,GV12}, one would expect that the accuracy of the Enskog result for the shear viscosity (which is the transport coefficient involved in the evaluation of the critical length scale) by considering only the leading term in a Sonine polynomial expansion \cite{GHD07} would decease as the mass ratio becomes more disparate. Thus, it is reasonably to expect that the second Sonine correction to $\eta$ mitigate part of the discrepancies observed in Ref.\ \cite{MGH14} for high dissipation and/or extreme mass or size ratios.

Before closing this Section, it is interesting to compare the predictions offered here with those obtained before \cite{JM89,Z95,AW98,AJ04} by neglecting the nonlinear dependence of the transport coefficients on dissipation. As mentioned in the Introduction, these works \cite{JM89,Z95,AW98,AJ04} are devoted to nearly elastic systems and hence, they assume the validity of energy equipartition. In fact, according to this level of approximation, the inelasticity in collisions is only accounted for in the cooling rate $\zeta_0$ and the expressions of the Navier-Stokes transport coefficients are the same as those obtained for elastic collisions \cite{LCK83}. In order to gauge the impact of the non-equipartition of granular energy and the $\al$-dependence of the transport coefficients, Fig.\ \ref{fig9} compares the dependence of $L_c$ on $m_1/m_2$ given by the present calculations (solid line) with those derived before (dashed line) \cite{JM89,Z95,AW98,AJ04}. In this plot, $x_1=0.1$, $\sigma_1/\sigma_2=1$, $\phi=0.1$ and $\al=0.7$. The comparison indicates good agreement between both approaches in the monodisperse gas case ($m_1=m_2$). On the other hand, quantitative significant differences appear as the mass ratio increases.

\section{Summary and discussion}
\label{sec5}

Particulate or granular flows play a critical role in chemical process industries (e.g., fluidized catalytic cracking), energy (e.g., gasification of coal and biomass), pharmaceuticals (e.g., powder processing, granulation), geological phenomena (e.g., planetary formation), and agriculture (e.g., grain conveying). In spite of this pervasiveness, granular systems are not completely understood yet. Apart from its practical interest, the study of granular matter under fluidization conditions poses open challenges from a fundamental point of view. In fact, granular matter can be considered as a good example of a \emph{complex} system since it is composed of many degrees of freedom (many particles) that are imbedded in a network of strong nonlinear interactions. This aspect explains in part because the use of non-equilibrium statistical physics or kinetic theory to describe granular fluids (i.e., when the material is externally excited) has been an active area of research in the past several decades. On the other hand, although in many conditions the motion of grains exhibits a great similarity to the random motion of atoms or molecules of an ordinary gas, the fact that collisions between grains are inelastic gives rise to subtle modifications of the conventional hydrodynamic equations.

In this paper a granular fluid mixture has been modeled as a mixture composed by smooth, inelastic hard spheres. The simplest situation for this system corresponds to the so-called homogeneous cooling state (HCS). It describes a uniform state with vanishing flow field and a granular temperature decreasing monotonically in time. However, it is well-known that the HCS can be instable against long wavelength spatial perturbations, leading to cluster and vortex formation. The existence of the instability can be predicted via a linear stability analysis of the nonlinear Navier-Stokes hydrodynamic equations. This analysis allows us to identify a critical length size $L_c$ beyond which the system becomes unstable. On the other hand, given that the expression of $L_c$ involves the Navier-Stokes transport coefficients of the system, one has to explicitly determine first these coefficients to get the impact of the parameter space (masses and sizes of grains, coefficients of restitution, composition, density, $\cdots$) on the critical size.

In order to gain some insight into the general problem, a kinetic theory description has been adopted where all the relevant information on the state of the system is given through the knowledge of the one-body distribution function for each species. For moderate densities, the Enskog equation \cite{GS95,BDS97} for smooth inelastic hard spheres can be considered as a reliable kinetic equation. As in the case of elastic collisions, the Enskog equation neglects velocity correlations between the particles which are about to collide (molecular chaos) but takes into account spatial correlations. To first order in the spatial gradients, the set of Enskog equations of the mixture have been solved \cite{GDH07,GHD07,MGH12} by means of the Chapman-Enskog method \cite{CC70} conveniently adapted to dissipative dynamics. As for ordinary fluid mixtures \cite{LCK83}, the Navier-Stokes transport coefficients are given in terms of a set of linear integral equations \cite{GDH07} that can be approximately solved by considering the leading terms \cite{GHD07,MGH12} in a Sonine polynomial expansion. The knowledge of the transport coefficients allows one to obtain the critical size for the onset of instability in terms of the parameters of the mixture. The present paper has addressed this problem for a $d$-dimensional granular binary mixture described by the Enskog equation. The results derived here cover some of the aspects not accounted for in previous studies. Specifically, (i) it takes into account the nonlinear dependence of transport on collisional dissipation (and thus, the theory is expected to hold for a wide range of coefficients of restitution); (ii) it considers the influence of energy non-equipartition on the critical size; and (iii) it is applicable to \emph{moderate} densities (say for instance, volume fraction typically smaller than or equal to 0.25 for hard spheres). Thus, the theory subsumes all previous studies for single fluids \cite{BDKS98,Mathieu1,Mathieu2,G05}, dilute mixtures \cite{GMD07,BR13} and nearly elastic dense mixtures \cite{JM89,Z95,AW98,AJ04}, which are recovered in the appropriate limits.

Our findings agree qualitatively well with previous results \cite{Mathieu2,JM89} carried out by using the \emph{elastic} expressions of the Navier-Stokes transport coefficients since the effect of dissipation on transport coefficients do not significantly change the form of the dispersion relations. Thus, stability analysis of the linearized hydrodynamic equations for the mixture shows $d-1$ transversal (shear) modes and one longitudinal heat mode. An analysis of the dependence of the above modes on the parameter space shows that in general the instability is driven by the transversal shear mode and hence, the critical size is given by Eq.\ \eqref{5.28.6} where the (reduced) cooling rate $\zeta_0^*$ is defined by Eq.\ \eqref{3.3} while the expressions defining the (reduced) shear viscosity $\eta^*$ are provided in the Appendix. It is quite apparent that $L_c/\sigma_{12}$ presents a complex dependence on the parameter space of the system so that, it is intricate to disentangle the impact of the different parameters on the critical size for instability. Thus, to reduce the number of independent parameters, a common coefficient of restitution ($\al_{ij}=\al$) has been assumed to illustrate the influence of the mass and size ratios as well as the composition on $L_c$. The results have been carried out in Section \ref{sec4} for hard disks ($d=2$) and spheres ($d=3$) in Figs.\ \ref{fig5}--\ref{fig8}. We observe that the role played by the mass and size ratios on the critical size is more relevant than that of the composition. In addition, although the dependence of $L_c$ on the parameter space is qualitatively similar for disks and spheres, the magnitude of the critical size is larger in the case of two dimensions than three dimensions for the same system.

With respect to the effect of collisional dissipation on $L_c$, as expected quantitative discrepancies between the present results and those obtained by assuming energy equipartition and elastic forms for transport coefficients \cite{JM89,Z95,AW98,AJ04} appear as the coefficient of restitution decreases. This is clearly illustrated in Fig.\ \ref{fig9}. Therefore, although the results derived in the quasielastic limit could predict reasonably well the dispersion relations, one expects that the results reported here improve these findings for conditions of practical interest where the transport coefficients are clearly affected by dissipation.

Although the present theory applies in principle for arbitrary values of the coefficients of restitution, it has some important restrictions. First, given that some previous computer simulation works \cite{SM01,SPM01,PTNE02} have clearly shown that the molecular chaos hypothesis fails for inelastic collisions as the density increases, it is possible that the limitations of the Enskog equation are greater than for elastic collisions. However, as mentioned in the Introduction, despite the above limitation the theoretical predictions obtained here from the Enskog equation compare well with recent MD simulations \cite{MGH14} of hard spheres. This shows again the reliability of the Enskog theory to accurately describe macroscopic properties (such as transport coefficients and/or the onset of instability) for a wide range of densities and/or coefficients of restitution. Another important limitation of
the theory is that the critical length obtained here has been estimated by considering only the first Sonine approximation for the shear viscosity coefficient $\eta$. Recent results \cite{GV09,GV12} for the tracer limit have shown that the reliability of the first Sonine solution can be questionable for strong collisional dissipation and/or disparate values of the mass and diameter ratios. Therefore, the second Sonine correction to $\eta$ could in part improve the agreement between theory and MD simulations \cite{MGH14} for $L_c$ when the mass ratio is large. The evaluation of the second Sonine approximation to the shear viscosity is an interesting open problem to be studied in the near future. Another possible direction of study is to consider the so-called modified Sonine method \cite{GSM07,GVM09} to determine $\eta$. This new approach consists of replacing, where appropriate in the Chapman-Enskog procedure, the MaxwellûBoltzmann distribution weight function (used in the standard first Sonine approximation) by the homogeneous cooling state distribution for each species. As in the case of dilute binary mixtures \cite{GVM09}, it is expected that the modified Sonine approximation improves the estimates of the standard one at strong dissipation. Work along these lines will be carried out in the near future.

\acknowledgments

The present research has been supported by the Spanish Government through grant No. FIS2013-42840-P, partially financed by
FEDER funds and by the Junta de Extremadura (Spain) through Grant No. GR15104.

\appendix
\section{Shear viscosity of a dense granular binary mixture}
\label{appA}

The explicit dependence of the (reduced) shear viscosity $\eta^*$ defined by Eq.\ \eqref{5.7} on the parameter space of the system is provided in this Appendix. The expression for $\eta^*$ reads
\beq
\label{a1}
\eta^*=\frac{4\pi^{(d-1)/2}}{\sqrt{2}(d+2)\Gamma\left(\frac{d}{2}\right)}\frac{\eta^{k*}+\eta^{c*}}{x_1\mu_{12}+x_2\mu_{21}}.
\eeq
The collisional contribution $\eta_c^*$ to the shear viscosity is \cite{GHD07,MGH12,GM03}
\beqa
\eta^{c*}&=&\frac{2\pi ^{d/2}}{\Gamma \left( \frac{d}{2}\right)}
\frac{n^*}{d(d+2)}\sum_{i=1}^{2}\sum_{j=1}^{2}x_{j}
\left(\frac{\sigma_{ij}}{\sigma_{12}}\right)^d\chi_{ij}\mu_{ji}
\nn
& \times & (1+\alpha _{ij})\eta_{i}^{k*}+\frac{d}{d+2}\kappa^{*},
\label{a2}
\eeqa
where the (reduced) bulk viscosity coefficient is
\beqa
\label{a3}
\kappa^{*}&=&\frac{4\pi ^{(d-1)/2}}{d^2\Gamma \left( \frac{d}{2}\right)}\frac{n^{*2}}{m_{1}+m_{2}}
\sum_{i=1}^{2}\sum_{j=1}^{2} x_i x_j \left(\frac{\sigma_{ij}}{\sigma_{12}}\right)^{d-1}
\nn
&\times& \chi_{ij} m_j \mu_{ij}
(1+\alpha _{ij})\left( \frac{\theta _{i}+\theta _{j}}{\theta_{i}\theta_{j}}\right)^{1/2}.
\eeqa
The partial kinetic contributions $\eta_{i}^{k*}$ read \cite{MGH12}
\begin{equation}
\label{a4}
\eta_{1}^{k*}=\frac{2(2\tau_{22}^*-c_d\zeta^*)\overline{\eta}_1-4\tau_{12}^*\overline{\eta}_2}
{c_d^2\zeta^{*2}-2c_d\zeta^*(\tau_{11}^*+\tau_{22}^*)+4(\tau_{11}^*\tau_{22}^*-\tau_{12}^*\tau_{21}^*)},
\end{equation}
\begin{equation}
\label{a5}
\eta_{2}^{k*}=\frac{2(2\tau_{11}^*-c_d\zeta^*)\overline{\eta}_2-4\tau_{21}^*\overline{\eta}_1}
{c_d^2\zeta^{*2}-2c_d\zeta^*(\tau_{11}^*+\tau_{22}^*)+4(\tau_{11}^*\tau_{22}^*-\tau_{12}^*\tau_{21}^*)},
\end{equation}
where $c_d\equiv [8\pi^{(d-1)/2}/(\sqrt{2}(d+2)\Gamma(d/2))]$, and
\beqa
\label{a6}
\overline{\eta}_i&=&x_i\gamma_i+\frac{\pi ^{d/2}}{d(d+2)\Gamma \left( \frac{d}{2} \right) }n^{*}
\sum_{j=1}^2
x_ix_j\left(\frac{\sigma_{ij}}{\sigma_{12}}\right)^d  \chi_{ij}\mu_{ji}
\nn
&\times & (1+\alpha _{ij})\left[
(3\alpha _{ij}-1)\left( \mu_{ji}\gamma_i+\mu_{ij}\gamma_{j}\right)
-4\mu_{ij}(\gamma_{i}-\gamma_{j})\right].\nn
\eeqa
The (reduced) collision frequencies $\tau_{ij}^*$ are given by \cite{GHD07,GM03}
%\begin{widetext}
\begin{eqnarray}
\label{a7}
\tau_{11}^*&=&\frac{2\pi^{(d-1)/2}}{d(d+2)\Gamma\left(\frac{d}{2}\right)}\left\{
x_1\left(\frac{\sigma_{1}}{\sigma_{12}}\right)^{d-1}(2\theta_1)^{-1/2}\right.\nn
& \times &(3+2d-3\alpha_{11})
(1+\alpha_{11})+2x_2 \mu_{21}(1+\alpha_{12}) \nn
& \times & \theta_1^{3/2}\theta_2^{-1/2}
\left[
(d+3)\left(\mu_{12}\theta_2-\mu_{21}\theta_1\right)\theta_1^{-2}(\theta_1+\theta_2)^{-1/2}
\right. \nn
& + & \frac{3+2d-3\alpha_{12}}{2}\mu_{21}
\theta_1^{-2}(\theta_1+\theta_2)^{1/2}\nn
&+& \left.\left.
\frac{2d(d+1)-4}{2(d-1)}\theta_1^{-1}(\theta_1+\theta_2)^{-1/2}\right]\right\},
\end{eqnarray}
\begin{eqnarray}
\label{a8}
\tau_{12}^*&=&\frac{4\pi^{(d-1)/2}}{d(d+2)\Gamma\left(\frac{d}{2}\right)}
x_1\chi_{12}\mu_{12}\theta_1^{-1/2}\theta_2^{3/2}
(1+\alpha_{12})\nonumber\\
& \times&\left[
(d+3)(\mu_{12}\theta_2-\mu_{21}\theta_1)\theta_2^{-2}(\theta_1+\theta_2)^{-1/2}\right.\nonumber\\
&+ &
\frac{3+2d-3\alpha_{12}}{2}\mu_{21}\theta_2^{-2}(\theta_1+\theta_2)^{1/2}
\nn
& -& \left. \frac{2d(d+1)-4}{2(d-1)}\theta_2^{-1}(\theta_1+\theta_2)^{-1/2}\right].
\end{eqnarray}
%\end{widetext}
The expressions for $\tau_{22}^*$ and $\tau_{21}^*$ can be obtained from Eqs.\ \eqref{a7} and \eqref{a8}, respectively, by just making the changes $1 \leftrightarrow 2$.

In order to get the dependence of the shear viscosity on the parameters of the system,
one needs to know the explicit forms of the pair correlations functions $\chi_{ij}$. For hard disks ($d=2$), a good approximation
for the pair correlation function $\chi_{ij}$ is \cite{JM87}
\begin{equation}
\label{a9}
\chi_{ij}=\frac{1}{1-\phi}+\frac{9}{16}\frac{\phi}{(1-\phi)^2}
\frac{\sigma_i\sigma_jM_1}{\sigma_{ij}M_2},
\end{equation}
where
\begin{equation}
\label{a10}
M_n=\sum_{s=1}^2\; x_s \sigma_s^n.
\end{equation}
In the case of hard spheres ($d=3$), we take for the pair correlation function
$\chi_{ij}$ the following approximation \cite{B70}
\begin{equation}
\label{a11}
\chi_{ij}=\frac{1}{1-\phi}+\frac{3}{2}\frac{\phi}{(1-\phi)^2}
\frac{\sigma_i\sigma_jM_2}{\sigma_{ij}M_3}
+\frac{1}{2}\frac{\phi^2}{(1-\phi)^3}\left(\frac{\sigma_i\sigma_jM_2}
{\sigma_{ij}M_3}\right)^2.
\end{equation}

\end{document}